\documentclass[conference, 10pt, a4paper]{IEEEtran}

\newtheorem{thm}{Theorem}

\newtheorem{lem}{Lemma}

\newtheorem{defn}{Definition}

\usepackage[final]{graphicx}
\usepackage[reqno]{amsmath}
\usepackage{amssymb}
\usepackage{subfig}
\usepackage{epstopdf}
\usepackage{bm}

\begin{document}

\sloppy

\title{Cloud-Based Topological Interference Management: A Case with No Cooperative Transmission Gain}
\author{\IEEEauthorblockN{Aly El Gamal\\}
 \IEEEauthorblockA{ECE Department, Purdue University\\ Email: \{elgamala\}@purdue.edu}}

\maketitle

\begin{abstract}
We study the problem of managing interference in linear networks, with backhaul constraints that admit centralized allocation of messages to transmitters through the cloud. Our setting is that of a generic channel, where no channel state information is available at the transmitters. Knowing only the network topology, we characterize the optimal decisions for assigning messages to transmitters, given that each receiver is interested in one message that can be available at $N$ transmitters. We show that using linear cooperation schemes, the per user degrees of freedom does not increase as we increase $N$ beyond unity. Hence, we conclude for the considered problem that cooperative transmission does not increase the degrees of freedom.
\end{abstract}

\section{Introduction}
In the past decade, there has been a rising practical significance for Ad-hoc wireless networks. Starting with the emergence of sensor networks and the increased importance and distribution of local wireless networks like WiFi networks, and now with the new paradigm of heterogeneous networks, there is an interest to understand Ad-hoc networks even for the design of cellular networks. Further, the expected future popularity of device networks under the umbrella of the Internet of Things (IoT) is drawing attention to research in the area of Ad-Hoc wireless networks. 

At the core of the wireless communication task, lies the problem of interference management, as the interference limitation stands as a bottleneck towards increasing the reliable rate of communication. While there has been tremendous effort over the past two decades to tackle the interference management problem from both theoretical and practical aspects, most of the results obtained cannot apply directly to recent Ad-hoc networks. Traditional results on the interference management problem are either tailored for fixed infrastructural cellular networks or rely on distributed approaches where cell association and transmission schedule decisions are based on local knowledge of the network. 

Most interference management solutions for Ad-hoc networks rely on local knowledge to make distributed decisions. Many advances in distributed transmission schemes have accompanied the rise of sensor networks in the past decade. The reason is that prior to the new technology of cloud computing, the common wisdom was that centralized solutions or relying on global network knowledge would not be feasible for networks that are not supported by fixed planned infrastructure. Further, the nature of cellular networks prior to the novel paradigm of heterogeneous networks did not include any form of Ad-hoc networks. Now, with these two new advances: Cloud computing and heterogeneous networks, a centralized approach for managing wireless transmission in Ad-hoc networks is both more feasible and enjoys significant practical relevance (see e.g.,~\cite{CRAN}-\cite{CRAN-Simeone-2}). 

We attempt to tackle the problem of managing interference in Ad-hoc networks in this work through studying the information-theoretic model of a linear interference network (introduced by Wyner in~\cite{Wyner}) with no channel state information available at the transmitters (no CSIT). Transmitters are only aware of the network topology in our model. Further, to capture the benefit of the cloud we make two assumptions. First, global information about the network topology is available. Second, each receiver is interested in one message that can be available at $N$ transmitters. The constraints on the number of transmitters that can be aware of each message relects a limited backhaul capacity constraint. Assigning messages to transmitters in this model mirrors taking cell association decisions in a cellular heterogeneous network. Further, the flexibility of being able to assign any message to any subset of $N$ transmitters reflects a cloud-based centralized controller that takes message assignment (cell association) as well as transmission schedule decisions. 

System models that share similarities with the considered model have been studied thoroughly in the literature. In particular, with the availability of channel state information at the transmitters, the problem of finding message assignments to transmitters that maximize the asymptotic per user degrees of freedom (DoF) has been solved for linear interference networks in~\cite{ElGamal-Annapureddy-Veeravalli-IT14}. It was shown in this work that combining centralized decisions for message assignment with cooperative transmission can achieve significant DoF gains through a simple delay-free zero-forcing transmit beamforming scheme. The question we answer in this work with regard to this previous work is whether the same insight hold even with no CSIT.

The problem of interference management with no CSIT, known as the topological interference management problem, has been studied with no cooperative transmission in~\cite{jafar-topological,Naderi-ElGamal-Avestimehr} and\cite{ElGamal-Naderi-Avestimehr}. In~\cite{jafar-topological} and~\cite{ElGamal-Naderi-Avestimehr}, this problem was considered for both constant channel and time-varying channel models, respectively. While the constant channel model assumes that the channel coefficients remain the same during a whole block of communication time slots, the time-varying channel model assumes a coherence time of unity, meaning that the channel changes from each time slot to the next. The results in these works show that DoF conclusions as well as the design of optimal interference management schemes differ dramatically between these two channel models. In this work, we show that enabling centralized decisions for assigning messages to transmitters allows us to reach the same conclusion for both time-varying and constant channel models, as long as the coherence time is the same for all communication links.

Recently, the problem of interference management through cooperative transmission has been studied with weak and no CSIT in~\cite{Gesbert-weakcsit}-\cite{Jafar-solution}. In~\cite{Gesbert-nocsit}, it was shown that assigning each message to all the transmitters connected to the desired receiver is beneficial compared to assigning each message only to the transmitter having the same index as the desired receiver. However, the proposed coding scheme in~\cite{Gesbert-nocsit} relies solely on interference avoidance and no cooperative transmission is exploited. In our problem, we try to understand the effect of cooperative transmission by allowing for a flexible assignment of messages to transmitters even when each message can be available at exactly one transmitter. 

The main conclusion of this work is that cooperative transmission does not increase the per user degrees of freedom in large linear interference networks. We reach this conclusion by proving an information theoretic converse for the case when each message can be available at more than one transmitter, i.e., $N > 1$. We prove that the asymptotic per user DoF for any $N > 1$ is the same as the one for the case when $N=1$, which is achieved by interference avoidance.

The remainder of this paper is organized as follows. In Section~\ref{sec:problem}, we introduce the problem formulation. The main result is stated in Section~\ref{sec:result}. We present the interference avoidance scheme for the case when $N=1$ in Section~\ref{sec:example}. We then prove the information theoretic upper bound for the case when $N>1$ in Section~\ref{sec:converse}. Before concluding, we discuss an extension of our problem to arbitrary network topologies in Section~\ref{sec:discussion}. We finally provide concluding remarks in Section~\ref{sec:conclusion}.

\section{Problem Formulation}\label{sec:problem}
We use the standard model for the $K-$user interference channel with a single antenna at each node.
\begin{equation}\label{eq:signal}
Y_i(t) = \sum_{j=1}^{K} H_{i,j}(t) X_j(t) + Z_i(t),i\in[K],
\end{equation}
where $t$ is the time index, $X_i(t)$ is the transmitted signal of transmitter $i$, $Y_i(t)$ is the received signal of receiver $i$, $Z_i(t)$ is the zero mean unit variance Gaussian noise at receiver $i$, $H_{i,j} (t)$ is the channel coefficient from transmitter $j$ to receiver $i$ over the $t^{th}$ time slot, and $[K]$ denotes the set $\{1,2,\ldots,K\}$. 

For any set ${\cal A} \subseteq [K]$, we define the complement set $\bar{\cal A} = \{i: i\in[K], i\notin {\cal A}\}$. For each $i \in [K]$, let $M_i$ be the message intended for receiver $i$, we use the abbreviations $X_{\cal A}$ and $Y_{\cal A}$ to denote the sets $\{X_i, i\in {\cal A}\}$ and $\{Y_i, i\in {\cal A}\}$, respectively.

\subsection{Channel Model}\label{sec:model}
Each transmitter is connected to its corresponding receiver as well as one following receiver, and the last transmitter is connected only to its corresponding receiver. More precisely,

\begin{equation}\label{eq:channel}
H_{i,j} \neq 0 \text { if and only if } i \in \{j,j+1\},\forall i,j \in [K].
\end{equation}

\begin{figure}[htb]
\centering
\includegraphics[width=0.8\columnwidth]{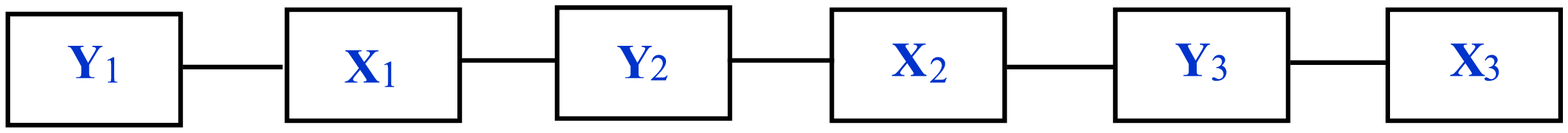}
\caption{Wyner's linear asymmetric model for $K=3$. In the figure, a solid line connects a transmitter-receiver pair if and only if the channel coefficient between them is not identically zero.}
\label{fig:wynermodel}
\end{figure}

The channel connectivity model is illustrated for $K=3$ in Figure~\ref{fig:wynermodel}.

Each non-zero channel coefficient is drawn independently from the same continuous distribution. Unless stated otherwise, all results in the paper are valid regardless of the coherence time of the channel (whether the channel remains constant across time slots or changes). While all receivers are assumed to be aware of the channel state information, the knowledge available for the design of the transmission scheme is that of the network topology. In other words, no channel state information is available at transmitters (no CSIT). 

\subsection{Cooperation Model}
For each $i \in [K]$, let ${\cal T}_i \subseteq [K]$ be the transmit set of message $M_i$. The transmitters in ${\cal T}_i$ cooperatively transmit the message $M_i$ to the receiver $i$. The messages $\{M_i\}$ are assumed to be independent of each other. The \emph{cooperation constraint} $N$ is defined as the maximum size of a transmit set:
\begin{equation}\label{eq:coop_order}
N = \max_i |{\cal T}_i|.
\end{equation}

\subsection{Linear Cooperation Schemes}
In this work, we restrict our attention to linear precoding schemes, where the transmit signal at each transmitter is given by a linear combination of signals; each depending only on one message. More precisely,
\begin{equation}
X_j = \sum_{i: j\in{\cal T}_i} X_{j,i}, \forall j\in[K],
\end{equation}
where $X_{j,i}$ depends only on message $M_i$. 

Each message $W_i$ is represented by a vector $\mathbf{w}_i\in\mathbb{C}^{m_i}$ of $m_i$ complex symbols that are desired to be delivered to the $i^{\text{th}}$ receiver. This message is encoded to one or both of the transmit vectors $\mathbf{X}_{i,i}^n=V_{i,i}^n \mathbf{w}_i$ and $\mathbf{X}_{i-1,i}^n=V_{i-1,i}^n \mathbf{w}_i$ , where $V_{i,i}^n$ and $V_{i-1,i}^n$ denote the $n\times m_i$ linear beamforming \emph{precoding} matrices used by transmitters $i$ and $i-1$ to transmit $W_i$. The rank of $V_{i,i}^n$ ($V_{i-1,i}^n$) is $m_{i,i}$ ($m_{i-1,i}$), where $m_{i,i} \leq m_i$ and $m_{i-1,i} \leq m_i$. Under such a scheme, the received signal of receiver $j$ over the $n$ time slots in (\ref{eq:signal}) can be rewritten as
\begin{eqnarray}
\mathbf{Y}_j^n&=&\left((H_{j,j-1}^n V_{j-1,j}^n) + (H_{j,j}^n V_{j,j}^n)\right) \mathbf{w}_j\nonumber\\&+&\sum_{(l,i): i\in[K], l \in \{j-1,j\} \cap {\cal T}_i} (H_{j,l}^n V_{l,i}^n) \mathbf{w}_i+\mathbf{z}_j^n,
\end{eqnarray}
where for every $i,j\in\{1,...,K\}$, $H_{i,j}^n$ is an $n\times n$ diagonal matrix with the $k^{\textrm{th}}$ diagonal element being equal to the value of the channel coefficient between transmitter $i$ and receiver $j$ in time slot $k$ . Each precoding matrix $V_{k,i}^n$ is an $n \times m_i$ matrix that can only depend on the knowledge of topology, and has rank $m_{k,i} \leq m_i, \forall k \in {\cal T}_i$. 

\subsection{Degrees of Freedom}
The total power constraint across all the users is $P$.  The rates $R_i(P) = \frac{\log|M_i|}{n}$ are achievable if the error probabilities of all messages can be simultaneously made arbitrarily small for a large enough block length $n$. The capacity region $\mathcal{C}(P)$ is the set of all achievable rate tuples. The DoF ($\eta$) is defined as $\limsup_{P \rightarrow \infty}\frac{ C_{\Sigma}(P)}{\log P}$, where $C_\Sigma(P)$ is the sum capacity. Since $\eta$ depends on the specific choice of transmit sets, we define $\eta(K,N)$ as the best achievable $\eta$ over all choices of transmit sets satisfying the cooperation order constraint in \eqref{eq:coop_order} for a $K-$user channel satisfying~\eqref{eq:channel}.  We define the per user DoF $\tau(N)$ to measure how the sum degrees of freedom scales with the number of users for a fixed cooperation order.
\begin{equation}
\tau(N) = \lim_{K\rightarrow \infty} \frac{\eta(K,N)}{K}
\end{equation}

It is worth noting here that modifying the channel model such that the channel coefficient between the last transmitter and first receiver ($H_{1K}$) is non-zero (cyclic model) does not change the value of $\tau(N)$.

\section{Main Result}\label{sec:result}
The main result in this work is a characterization of the asymptotic per user DoF $\tau(N)$ for all values of the cooperation constraint $N$ in Wyner's linear interference networks.
\begin{thm}\label{thm:main}
Under the restriction to linear cooperation schemes, transmitter cooperation with no CSIT does not increase the asymptotic per user DoF in linear interference networks. More precisely,
\begin{equation}
\tau(N)=\tau(1)=\frac{2}{3}, \forall N \in {\bf Z}^+.
\end{equation}
\end{thm}

\section{Achieving $\frac{2}{3}$ per user DoF without Cooperation}\label{sec:example}
Consider the following message assignment for the case where $N=1$. The problem in this case is a link scheduling problem whose solution is given as follows.

\[
{\cal T}_i = \begin{cases}
\{i\}, & i \text{ mod } 3 = 1,\\
\{i-1\}, & i \text{ mod } 3=0.
\end{cases}
\]

Further, the messages $M_i, i \text{ mod } 3 = 2$ are not transmitted. Recall that the objective here is to maximize the sum degrees of freedom, so deactivating some transmitters or not transmitting some messages can make sense. Now, we can see that message $M_i, i \text{ mod } 3=1$ can be delivered without interference through $X_i$ to $Y_i$. Also, message $M_i, i \text{ mod } 3=0$ can be delivered without interference through $X_{i-1}$ to $Y_i$. For every three messages, two are delivered to their intended receivers without interference and one is not transmitted. Hence, $2$ degrees of freedom are achieved for each group of successive $3$ messages.

\section{Converse Proof}\label{sec:converse}
In order to prove the converse, we prove a statement about general network topologies. First, we make that statement and show how it implies our desired upper bound, and then we prove it. We define the following condition to refer to a class of network topologies of interest. For any network topology, let ${\cal N}({\cal A})$ be the set of indices of transmitters connected to at least one receiver with an index in ${\cal A}$. In other words, ${\cal N}({\cal A})$ are the neighbors of receivers in ${\cal A}$.
\begin{defn}
We say that a network topology satisfies Condition $1$ if there exists a set ${\cal A}=\{a_1,\cdots,a_{|{\cal A}|}\} \subseteq [K]$ of receiver indices such that the neighboring sets $\{{\cal N}\left(\{i\}\right):i\in{\cal A}\}$ do not overlap. Further, if we remove all receivers with indices in ${\cal A}$, as well as all transmitters with indices in $\bar{\cal N}({\cal A})$ from the network, the remaining bipartite graph has a matching that covers each transmitter in ${\cal N}({\cal A})$.
\end{defn}
\begin{lem}\label{lem:one}
For any network topology satisfying Condition $1$, the sum DoF $\eta \leq |{\bar{\cal A}}|$.  
\end{lem}

\begin{figure}[htb]
\centering
\includegraphics[width=0.3\columnwidth]{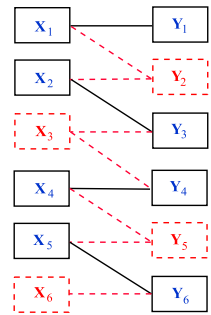}
\caption{Wyner's linear asymmetric model satisfies Condition $1$ with ${\cal A}=\{i:i \text{ mod } 3=2\}$. After removing the dashed red nodes and edges, we end up with a perfect matching.}
\label{fig:Condition1Fig}
\end{figure}

For the linear interference network topology considered in this work, Condition $1$ is satisfied for any network with a number of users $K$ such that $K \text{ mod } 3 = 0$, with a set ${\cal A}=\{i:i \text{ mod } 3=2\}$ (see Figure~\ref{fig:Condition1Fig}). Hence, applying Lemma~\ref{lem:one} for large networks, we reach the result that $\tau(N) \leq \frac{2}{3}$, regardless of the value of the cooperation constraint $N$.

We now dedicate the rest of this section to prove Lemma~\ref{lem:one}. We provide the proof here for linear interference networks, and delegate the general proof of any network topology satisfying Condition $1$ to the journal version of this work. The first step is to prove the following lemma about reconstructing a message from the knowledge of signals observed at the receivers it interferes at.

\begin{lem}\label{lem:noiccanc}

For $i \geq 3$, if $[H_{i,i}^n V_{i,i}^n + H_{i,i-1}^n V_{i-1,i}^n]$ has full column rank of $m_i$ almost surely, then,
\begin{eqnarray}
 &&\mathsf{rank}([I_{i,i-1}^n \quad I_{i,i+1}^n])\nonumber\\&&\overset{a.s.}{\geq} \mathsf{rank}\left([H_{i,i}^n V_{i,i}^n +H_{i,i-1}^n V_{i-1,i}^n]\right)\overset{a.s.}{\geq} m_i,
\end{eqnarray}
where $I_{i,i-1}^n$ and $I_{i,i+1}^n$ represent the interference caused by message $W_i$ at receivers $i-1$ and $i+1$, respectively, over $n$ time slot. More precisely, $I_{i,i-1}^n=[H_{i-1,i-1}^n V_{i-1,i}^n + H_{i-1,i-2}^n V_{i-2,i}^n]$ and $I_{i,i+1}^n=[H_{i+1,i}^n V_{i,i}^n +H_{i+1,i+1}^n V_{i+1,i}^n]$.
\end{lem}
\begin{IEEEproof}
We provide the main concepts behind the proof. Note that,
\begin{eqnarray}\label{eq:icdecomposition}
[I_{i,i-1}^n \quad I_{i,i+1}^n]&=&[H_{i-1,i-1}^n V_{i-1,i}^n \quad H_{i+1,i}^n V_{i,i}^n]\nonumber\\&+&[H_{i-1,i-2}^n V_{i-2,i}^n \quad H_{i+1,i+1}^n V_{i+1,i}^n].
\end{eqnarray}
Further,
\begin{eqnarray}
&&\mathsf{rank}([H_{i-1,i-1}^n V_{i-1,i}^n \quad H_{i+1,i}^n V_{i,i}^n])\nonumber\\&\geq& \mathsf{rank}([H_{i-1,i-1}^n V_{i-1,i}^n+H_{i+1,i}^n V_{i,i}^n]) \overset{a.s.}{\geq} m_i. 
\end{eqnarray}
Finally, we prove by induction on the parameter $n$ that the addition of $[H_{i-1,i-2}^n V_{i-2,i}^n \quad H_{i+1,i+1}^n V_{i+1,i}^n]$ in~\eqref{eq:icdecomposition} does not reduce the rank of the sum almost surely. In order to do so, we first define the following,
\begin{eqnarray}
{\bf M}_1^n&=& [H_{i-1,i-1}^n V_{i-1,i}^n \quad H_{i+1,i}^n V_{i,i}^n],\\
{\bf M}_2^n&=& [H_{i-1,i-2}^n V_{i-2,i}^n \quad H_{i+1,i+1}^n V_{i+1,i}^n].
\end{eqnarray} 
For the base of the induction, we consider the case where $n=1$. In this case, ${\bf M}_1^n$ is just one row of $2m_i$ elements. Since $H_{i-1,i-2}^n$ and $H_{i+1,i+1}^n$ are drawn from a continuous distribution and independently from $H_{i-1,i-1}^n$ and $H_{i+1,i}^n$, the probability that ${\bf M}_2^{n=1} = -{\bf M}_1^{n=1}$ is zero, and hence, rank$\left({\bf M}_1^{n=1}+{\bf M}_2^{n=1}\right)$ $\geq$ rank$\left({\bf M}_1^{n=1}\right)$ almost surely.

For the induction step, we assume that rank$\left({\bf M}_1^{n}+{\bf M}_2^{n}\right)$ $\geq$ rank$\left({\bf M}_1^{n}\right)$ almost surely, and want to prove that for $m_i \geq n+1$,

\begin{equation}\label{eq:step}
\text{rank}\left({\bf M}_1^{n+1}+{\bf M}_2^{n+1}\right) \overset{a.s.}{\geq} \text{rank}\left({\bf M}_1^{n+1}\right) 
\end{equation} 

Let ${\bf M}_1^{n+1,1:n}$ and ${\bf M}_2^{n+1,1:n}$ be the submatrices of ${\bf M}_1^{n+1}$ and ${\bf M}_2^{n+1}$ that consist of the first $n$ rows. We know from the induction hypothesis that rank$\left({\bf M}_1^{n+1,1:n}+{\bf M}_2^{n+1,1:n}\right)$ $\geq$ rank$\left({\bf M}_1^{n+1,1:n}\right)$ almost surely. Hence, if~\eqref{eq:step} does not hold, then it has to be the case that the last row of ${\bf M}_1^{n+1}$ is not in the row span of the first $n$ rows, with a non-zero probability. Further, the last row of ${\bf M}_1^{n+1}+{\bf M}_2^{n+1}$ has to be in the row span of the first $n$ rows, with a non-zero probability. Here, we reach a contradiction becasue $m_i \geq n+1$ and the channel coefficients drawn in the ${n+1}^{\text{st}}$ time slot are independent from all previous channel coefficients.
\end{IEEEproof}
We now restate \cite[Lemma $4$]{ElGamal-Annapureddy-Veeravalli-IT14} without proof.
\begin{lem}\label{lem:dofouterbound}
If there exists a set ${\cal B}\subseteq [K]$, a function $f_1$, and a function $f_2$ whose definition does not depend on the transmit power constraint $P$, and $f_1\left(Y_{\cal B},X_{U_{\cal B}}\right)=X_{\bar{U}_{\cal B}}+f_2(Z_{\cal B})$, then the sum DoF $\eta \leq |{\cal B}|$. 
\end{lem} 
In the above lemma, we used $U_{\cal B}$ as the set of indices of transmitters that exclusively carry the messages for the receivers in ${\cal B}$, and the complement set $\bar{U}_{\cal B}$ is the set of indices of transmitters that carry messages for receivers outside ${\cal B}$. More precisely, $U_{\cal B} = [K]\backslash\cup_{i \notin {\cal B}} {\cal T}_i$. 

Now, in order to prove Lemma~\ref{lem:one} for linear interference networks, we use Lemma~\ref{lem:dofouterbound} with ${\cal B}=\bar{\cal A}=\{i \in [K]: i \text{ mod } 3 \neq 2\}$. The set of transmit signals $X_{\bar{U}_{\cal B}}$ is the set $\{X_{j,i}, i\in {\cal A}, j\in{\cal T}_i\}$. Note that we can reconstruct $X_{\bar{U}_{\cal B}}$ if we can reconstruct the symbols $\{{\bf w}_i: i \in {\cal A}\}$. Assume w.l.o.g. that $K \text{ mod } 3 = 0$. We now apply Lemma~\ref{lem:dofouterbound} by designing the functions $f_1$ and $f_2$ such that we remove the contribution of the signals $X_{U_{\cal B}}$ and $Z_{\cal B}$ from $Y_{\cal B}$, to obtain the matrices $\{I_{i,i-1}^n,I_{i,i+1}^n: i \in {\cal A}\}$ that are used in Lemma~\ref{lem:noiccanc}. From these matrices, we can reconstruct the desired symbols almost surely, and hence, the statement of Lemma~\ref{lem:dofouterbound} holds. 

The general version of this argument would use the fact that neighboring sets of receivers in ${\cal A}$ do not overlap, $|X_{\bar{U}_{\cal B}}| \leq |{\cal N}\left({\cal A}\right)|$, and then we obtain $|\bar{\cal A}|$ equations in at most $|{\cal N}\left({\cal A}\right)|$ variables. From Condition $1$, we know that each variable appears in at least one unique equation (for almost all channel realizations), and hence the system of linear equations can be solved to obtain the signals $X_{\bar{U}_{\cal B}}$

%

\section{Discussion: Arbitrary Network Topologies}\label{sec:discussion}
\subsection{Receivers with Identical Neighboring Sets}\label{sec:identicalneighbors}
There is a feature of linear interference networks that enables the proof of Theorem~\ref{thm:main}. When we look at the set ${\cal A}=\{i:i \text{ mod } 3 = 2\}$ that is used to prove that linear interference networks satisfy Condition $1$, we note that the size of the complement set $|\bar{\cal A}|=|{\cal N}\left({\cal A}\right)|$.  In fact, for Condition $1$ to be satisfied, a necessary condition would be that $|\bar{\cal A}| \geq |{\cal N}\left({\cal A}\right)|$. Now, let's consider an extreme example where this necessary condition is violated for all choices of the receiver set ${\cal A}$. Consider a network where the following holds,
\begin{equation}\label{eq:violation}
\forall {\cal A}\subseteq [K]: |{\cal A}| \geq 1, |\bar{\cal A}| < |{\cal N}({\cal A})|.
\end{equation}
The network satisfying~\eqref{eq:violation} is the fully connected network. Hence, using the same technique for proving a converse as in Section~\ref{sec:converse} would not yield a low upper bound on the DoF for fully connected networks. However, one can use a different argument for this network: All receivers in a fully connected network observe a linear combination of all transmit signals, but the coefficients of these linear combinations vary because of the different channel coefficients. However, since we are assuming that the channel coefficients are not known at transmitters and each is drawn independently from the same distribution, the following holds. If receiver $1$ can decode message $M_1$ for almost all channel realizations, then any other receiver will also be able to decode $M_1$ for almost all channel realizations, regardless of the choice of message assignment and coding scheme. Further, the same conclusion holds for any other message $M_i, i\in\{2,\cdots,K\}$. It follows that the sum DoF is bounded by the sum DoF of a network that has $K$ transmitters and only one receiver, which is unity. Hence, transmitter cooperation cannot increase the DoF in fully connected networks as well, even if we are not restricted to linear cooperation schemes.

We can generalize the above argument for fully connected networks to any general topology. That is, whenever there are two or more receivers connected to the same set of transmitters, the sum DoF for messages intended at these receivers is unity. We haven't used this fact in Section~\ref{sec:converse} because for linear interference networks, each receiver is connected to a unique set of transmitters.


\subsection{Potential Cooperative Transmission Gains}
\begin{figure}[htb]
\centering
\includegraphics[width=0.3\columnwidth]{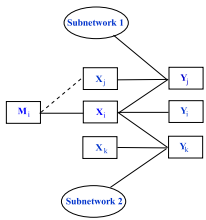}
\caption{Example with potential cooperative transmission gain}
\label{fig:Case1Fig}
\end{figure}

Another way to look at our result for linear interference networks is that it is not useful to assign any message $M_i$ to a transmitter that is not connected to receiver $i$. We do not know whether this conclusion holds for general network topologies. In particular, consider a network topology as the one depicted in Figure~\ref{fig:Case1Fig}, where subnetworks $1$ and $2$ represent arbitrary sets of transmitters. Message $M_i$ is delivered in the first time slot to receiver $i$ through $X_i$. Now, suppose that in the second time slot, we want to repeat the transmission of $M_i$ to cancel its interference at $Y_j$, without causing interference at $Y_k$ because another transmitter in subnetwork $2$ is delivering a message to receiver $k$ in the second time slot. The only way to do this is to transmit $M_i$ from $X_j$ in the second time slot. However, we do not know whether this scenario can appear in an optimal coding scheme. We just provide this example to stimulate thinking about solving the problem for general network topologies.

\subsection{Why Identical Channel Distributions}~\label{sec:identical}
While we have not made an assumption on the coherence time of the channel in Section~\ref{sec:model}, we have assumed that all channel coefficients are drawn from the same distribution. This assumption was essential to our converse proof in Section~\ref{sec:converse} as well as the argument we made in Section~\ref{sec:identicalneighbors}. In this section, we consider an example where assuming that channel coherence time varies across channels that are observed at different receivers, leads to a different conclusion. 

\begin{figure}[htb]
\centering
\includegraphics[width=0.3\columnwidth]{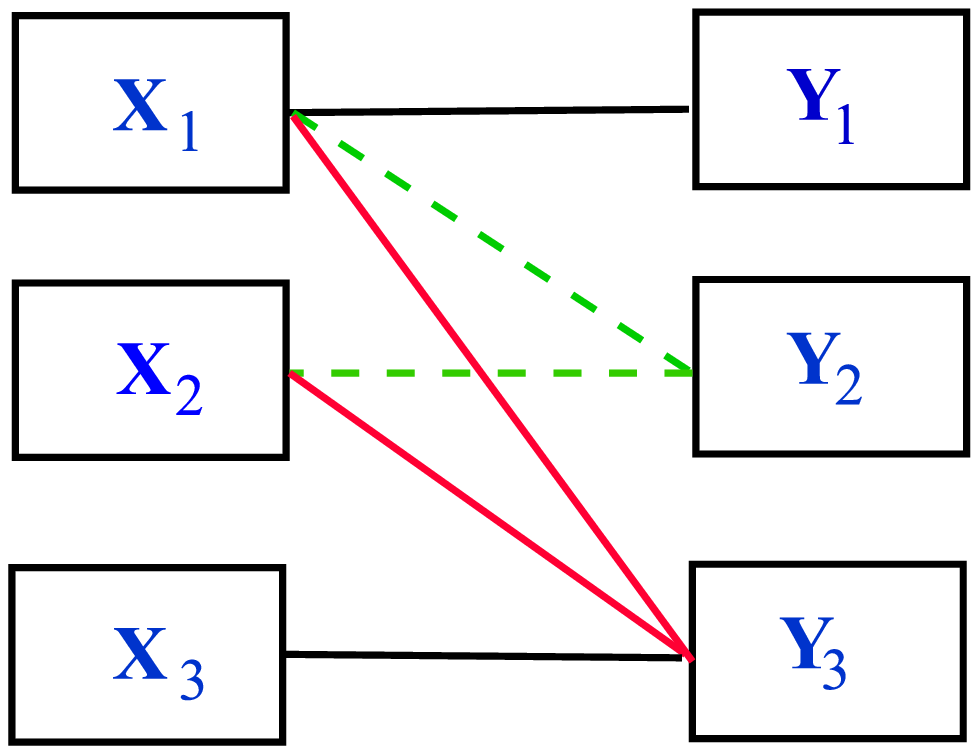}
\caption{Example channel model discussed in Section~\ref{sec:identical}}
\label{fig:Case2Fig}
\end{figure}

Consider the example $3$-user network in Figure~\ref{fig:Case2Fig}, and assume that $M_i$ is only available at transmitter $i$. i.e., ${\cal T}_i=\{i\}, \forall i \in \{1,2,3\}$. If all channel coefficients are drawn from the same distribution, then the sum DoF is unity. The converse follows from the following argument. Assuming a reliable communication scheme, if receiver $3$ can decode $M_3$, then it can remove the contribution of $X_3$ from $Y_3$, and consequently, obtain a statistically equivalent signal to $Y_2$. Hence, if receiver $2$ can decode $M_2$, then so will receiver $3$. Receiver $3$ can then remove the contribution of $X_2$ from $Y_3$, and finally obtain a statistically equivalent signal to $Y_1$. It follows that all messages can be decoded at receiver $3$, and hence, the sum DoF is at most unity. 

If we consider the scenario where the channels $H_{2,1}$ and $H_{2,2}$ (dashed green in Figure~\ref{fig:Case2Fig}) have coherence time of unity but $H_{3,1}$ and $H_{3,2}$ (solid red in Figure~\ref{fig:Case2Fig}) have coherence time of two, then $\frac{3}{2}$ DoF is achievable. The achievability scheme is based on achieving $3$ DoF in $2$ time slots. In the first time slot, all three transmit signals are active, and in the second time slot, transmitters $1$ and $2$ repeat their transmissions. Receiver $2$ then obtains two equations in $X_1$ and $X_2$, and these two equations are linearly independent almost surely. Hence, receiver $2$ can decode both $X_1$ and $X_2$. Since $H_{3,1}$ and $H_{3,2}$ remain constant over the two time slots, receiver $3$ can use its received signal in the second time slot, to cancel the interference received in the first time slot. It follows that all receivers can decode their desired messages over two time slots, and hence, $\frac{3}{2}$ DoF is achieved. 


\section{Conclusion}\label{sec:conclusion}
We characterized the asymptotic per user DoF for linear interference networks with no CSIT. We considered the possibility of transmitter cooperation by allowing for assigning each message to $N$ transmitters. Under the restriction to linear cooperation schemes, we proved that transmitter cooperation is not useful in the considered setting. To prove the result, we showed that the per user DoF for any value of $N$ can be achieved by assigning each message to a single transmitter and using an interference avoidance scheme.

We then discussed in Section~\ref{sec:discussion} the extension of our result to general network topologies. We argued that the fully connected network admits the same conclusion about the utility of transmitter cooperation, even without the restriction to linear cooperation schemes. We then shed light on the difficulty of making the same claim to any network topology through an exemplary scenario. Finally, we showed through an example why changing the assumption of identical coherence time for all channel coefficients may lead to different conclusions and insights.

We do not know whether transmitter cooperation can be useful, from a DoF point of view, in any network topology where no channel state information is available at any transmitter. We are interested in attempting to answer this question in future work.

\bibliographystyle{IEEEtran}

\begin{thebibliography}{15}
\bibitem{CRAN}
S.~Veetil, K.~Kuchi and R.~K.~Ganti. (2015, Dec.). Performance of cloud radio access networks. [Online]. Available: http://arxiv.org/pdf/1512.05904v1.pdf

\bibitem{CRAN-2}
A.~Checko, H.~L.~Christiansen, Y.~Yan, L.~Scolari, G.~Kardaras, M.~S.~Berger and L.~Dittmann, ``Cloud RAN for mobile networks - a technology overview," \emph{IEEE Communication Surveys Tutorials}, vol.~17, no.~1, pp.~405-426, First Quart. 2015.

\bibitem{CRAN-3}
China Mobile, ``Next generation fronthaul interface," White Paper, Oct. 2015.

\bibitem{CRAN-4}
The 5G Infrastructure Public Private Partnership. (2015, Jan.). 5G-Xhaul Project. [Online]. Available: https://5g-ppp.eu/5g-xhaul/

\bibitem{CRAN-Simeone}
O.~Simeone, A.~Maeder, M.~Peng, O.~Sahin and W.~Yu. (2015, Dec.). Cloud radio access network: Virtualizing wireless access for dense heterogeneous systems. [Online]. Available: http://arxiv.org/abs/1512.07743.

\bibitem{CRAN-Simeone-2}
S.~-H.~Park, O.~Simeone and S.~Shamai. (2016, Jan.). Joint optimization of cloud and edge processing for fog radio access networks. [Online]. Available: http://arxiv.org/abs/1601.02460.

\bibitem{Wyner}
A.~Wyner, ``{S}hannon-theoretic approach to
  a {G}aussian cellular multiple-access channel,'' \emph{IEEE Trans.
  Inf. Theory}, vol.~40, no.~5, pp. 1713-1727, Nov. 1994.

\bibitem{ElGamal-Annapureddy-Veeravalli-IT14}
{A.~El Gamal}, V.~S.~Annapureddy, and V.~V.~Veeravalli, ``Interference channels with coordinated multi-point transmission: Degrees of freedom, message assignment, and fractional reuse", \emph{IEEE Trans. Inf. Theory}. vol.~60, no.~6, pp. 3483-3498, Mar. 2014.

\bibitem{jafar-topological}
S.~A.~Jafar, ``Topological interference management through index coding", \emph{IEEE Transactions on Information Theory}, vol.~60, no.~1, pp.~529-568, Jan. 2014.

\bibitem{Naderi-ElGamal-Avestimehr}
N.~Naderi, A.~{El Gamal} and S.~Avestimehr, ``Topological interference management with retransmission: What are the best topologies?," \emph{IEEE International Conference on Communications (ICC)}, London, Jun. 2015.

\bibitem{ElGamal-Naderi-Avestimehr}
A.~{El Gamal}, N.~Naderi and S.~Avestimehr, ``When does an ensemble of matrices with randomly scaled rows lose rank," \emph{IEEE International Symposium on Information Theory (ISIT)}, Barcelona, Jul. 2015.

\bibitem{Gesbert-weakcsit}
P.~{de Kerret} and D.~Gesbert. (2016, Feb.). Network MIMO: Transmitters with no CSI can still be very useful. [Online]. Available: http://arxiv.org/abs/1601.07399

\bibitem{Gesbert-2}
P.~{de Kerret} and D.~Gesbert, ``Degrees of freedom of the network MIMO channel with distributed CSI," \emph{IEEE Transactions on Information Theory}, vol.~58, no.~11, pp.~6806-6824, Nov. 2012.

\bibitem{Gesbert-nocsit}
X.~Yi and D.~Gesbert, ``Topological interference management with transmitter cooperation," \emph{IEEE International Symposium on Information Theory (ISIT)}, Honolulu, Jul. 2014.

\bibitem{Jafar-Goldsmith-2005}
S.~A.~Jafar and A.~J~Goldsmith, ``Isotropic fading vector broadcast channels: The scalar upper bound and loss in degrees of freedom," \emph{IEEE Transactions on Information Theory}, vol.~51, no.~3, pp.~848-857, Mar. 2005.

\bibitem{lapidoth-conjecture}
M.~Wigger, A.~Lapidoth and S.~Shamai. (2006.). On the capacity of fading MIMO broadcast channels with imperfect transmitter side information. [Online]. Available: http://arxiv.org/abs/cs/0605079.

\bibitem{Jafar-solution}
A.~G.~Davoodi and S.~A.~Jafar. (2014.). Aligned image sets under channel uncertainty: Setting a conjecture by lapidoth, shamai and wigger on the collapse of degrees of freedom under finite precision CSIT. [Online]. Available: http://arxiv.org/abs/1403.1541.

\end{thebibliography}

\end{document}